# ¿Es el fútbol una gran mentira o es simplemente un sistema complejo?

Nelson Fernández[1-3] y Ricardo Bernal[2-3]

[1]Lab. Sistemas Complejos, Universidad de Pamplona, Colombia. nfernandez@unipamplona.edu.co
[2]LigaMx. joseberna@ligamx.net
[3]CxSportsNet  https://sites.google.com/view/complexity-and-sports/cxsports-satellite?authuser=0

Entrenadores sin experiencia que alcanzan triunfos inesperados, equipos favoritos con grandes plantillas que pierden de manera estrepitosa, métodos rebatidos por la ciencia del deporte que los equipos implementan y generan extraños pero buenos resultados. El reconocimiento de situaciones como estas llevó a Pep Guardiola a considerar que: "El fútbol es una gran mentira". Su expresión corporal manifestaba gran sorpresa cuando al tiempo se preguntaba "¿cómo podría valer la experiencia si cada situación es nueva, sí la experiencia que te sirvió ayer no te sirve hoy?". Lo sustentaba con la anécdota de la final de la Champions 2004-2005 entre Milán y Liverpool, conocida como "El Milagro de Estambul". En esta final, el Milán tenía el equipo más experimentado que él hubiese visto (Cafú, Maldini, Dida, Iniesta, Gattuso, Pirlo, Crespo, Inzaghi, por mencionar algunos), guerreros de mil batallas. Se fueron al descanso ganando 3-0, y en el segundo, en seis minutos, recibieron los tres goles del empate. La definición por penales quedó 3-2 y ganó el Liverpool. ¿Qué pudo suceder? ¿La ilusión de los jóvenes del Liverpool le ganó a la experiencia del mejor equipo europeo del momento, y le dieron al Liverpool lo que necesitaba? Así de simple, dice Pep: funcionó esta vez; otra vez podría no funcionar, a pesar de ser los mismos con los mismos. ¿Por qué? Porque el fútbol además es un misterio; cualquiera que lo analice llegaría a la conclusión que el fútbol, siendo así, no es así.

Cuando aparecen este tipo de *misterios futbolísticos*, surge la ciencia como medio para dar explicaciones; y lo primero es basarnos en la naturaleza y los comportamientos colectivos de organismos que "hacen cálculos" para generar soluciones sorprendentes. Esto es parte de todo lo genial que surge al estudiar estos "Sistemas Complejos", a los que por definición no escapa el fútbol, y para el que podemos buscar más de una explicación bioinspirada.

No obstante, ante la variedad de interpretaciones que puede tener la complejidad, sus propiedades asociadas y lo que es un sistema complejo, se hace conveniente brindar algunos elementos para entender *como la impredecibilidad en el fútbol da paso a cientos de resultados contraintuitivos y como las ciencias de la complejidad podrían aportar al entendimiento de muchos fenómenos en este deporte*. Este es el objetivo que se persigue al abordar de manera sintética algunos de los aspectos más importantes de lo que sería la complejidad aplicada, y que discutimos en las siguientes secciones con el fin de acercar más la ciencia y el deporte.

# Contenido

# 1. Enfoques científicos para estudiar el fútbol: la reducción del sistema a las partes

Desde una base física podríamos describir al fútbol como un *sistema dinámico*; es decir, el fútbol en toda su dimensión y esencia cambia en el tiempo. "El chiste" sería hallar la regla de cambio o evolución para describirlo y poder generar algunas predicciones. Escogeríamos entre reglas determinísticas, probabilísticas o algorítmicas para comenzar a intentarlo. Sin embargo, al considerar anécdotas como la del milagro de Estambul, nos damos cuenta que "El fútbol, como el mundo, no es predecible". En este aspecto, hallar la "regla de cambio" que lo explique sería un tema pendiente.

Desde un enfoque científico tradicional o reduccionista, podríamos optar por separar y aislar los elementos del sistema futbolístico. Luego podríamos dividirlo en partes aún más pequeñas en busca de desentrañar su naturaleza más íntima. Con esto, generaríamos un conocimiento profundo en temas muy detallados, y se podría intentar volver a la predicción. De hecho, es común estudiar los distintos componentes del juego como los propuestos en la figura 1, en la que lo técnico- táctico, el rendimiento, la formación, el desarrollo, la planeación y la investigación, entre otros, pueden tomar lugar.

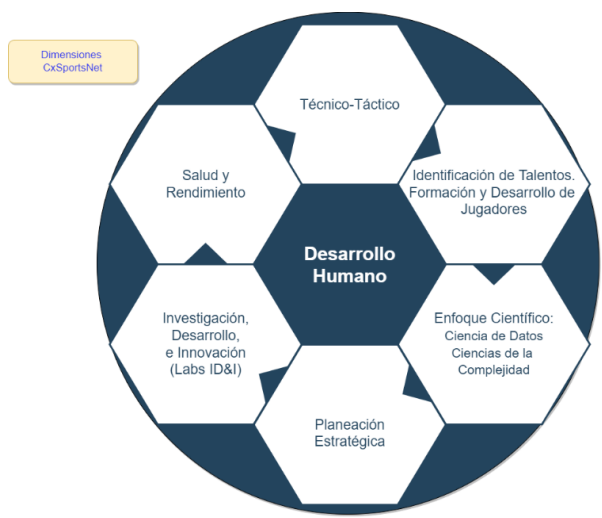

*Figura 1 Dimensiones de Desarrollo humano en el fútbol*

El avance del conocimiento en cada aspecto nos muestra que es común hallar a uno o más especialistas en las distintas temáticas o incluso a súper-especialistas que estudian algunos de estos componentes.

El problema de la especialización y subsecuente súper-especialización es que, cuando queremos volver a ensamblar el sistema como un todo, no podemos hacerlo porque hemos perdido de vista lo que lo mantenía unido, los conectores y sus relaciones. En consecuencia, tenemos limitaciones para observar la estructura del sistema (elementos y sus relaciones) y su orden u organización global. Este es el costo de comprender el sistema por partes, perdiendo de vista su unicidad, y dado que en el fútbol no opera el principio de superposición que define que el todo es igual a la suma de sus partes, no podremos hallar una solución "integral". Cabe destacar que el fútbol no es aditivo o sumativo, no es lineal.

Ante la separación que hace la ciencia al estudiar las partes, surge el reto de hallar alternativas, enfoques y métodos para estudiar también lo que las conecta, es decir los conectores y las relaciones en el sistema que es lo que lo mantiene unido.

## 2. La integración de los componentes: El fútbol como sistema complejo

El punto clave hasta acá, es que, el fútbol no se puede separar en componentes, desde el aislamiento de sus elementos, para explicarlo. Principalmente, porque entre sus elementos existen interacciones destacadas (*relevantes*) que no lo permiten. Sobre la base de este principio, es que comprendemos que el fútbol es un fenómeno y sistema complejo. Surge entonces la necesidad de definir lo que es Complejidad, y para ello nos basamos en una de las definiciones más importantes y que resulta de su etimología: *Complejidad* viene del latín *Plexus* que significa "Entretejido" (C. Gershenson & Fernández, 2012). Así, un sistema complejo es un sistema entrelazado difícil de separar por el entramado de sus interacciones relevantes que no permite estudiarlo de manera aislada.

Las interacciones relevantes se generan autónomamente, sin que haya un controlador de las mismas, evento que se denomina *autoorganización*. En este momento, se genera una *estructura* con un *orden* determinado que se va ajustando según se vayan dando más interacciones, por lo que muchas veces podemos hallar ciertos patrones que *emergen* en el comportamiento o la dinámica del sistema. Sin embargo, vemos cómo el sistema se adapta dinámicamente a los cambios del ambiente sin perder su autonomía.

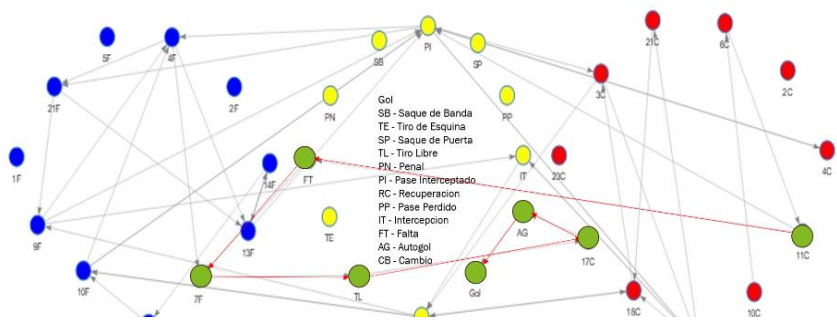

*Figura 2 Rede de Dinámica Acoplada de Eventos de Defensa y Ataque en la Final de Mundial de Fútbol 2018*

En el fútbol, la *estructura* de juego de un equipo se puede representar con redes de pases, o de manera más novedosa, redes de eventos, bien sea para un partido o una secuencia de ellos en un campeonato. Algo así como la *Red de Defensa y Ataque* (Redes DA, Fig. 2) que resultó de observar la final del Mundial 2018 entre Francia (red azul) y Croacia (red roja), basada en los eventos (red amarilla) que se dieron en el partido y donde se hizo posible observar la ruta de eventos (nodos verdes y líneas rojas) que antecedieron a uno de los goles (Fernández et al., 2018).

Si bien podemos hallar la estructura del juego y observar su orden resultante, a nivel de interacciones es claro que en el fútbol no sólo se dan interacciones en los 90 minutos de juego, sino que fuera de la cancha también las hay y entre muchos más actores.

Desde la perspectiva de la filosofía y modelo de juego que los clubes implementan, el fenómeno de autoorganización también se puede describir dentro de una subcategoría: la "autoorganización guiada". El planteamiento y la planeación táctica del partido tiene reglas definidas que el entrenador espera que los jugadores cumplan en el partido de ahí la guía para alcanzar lo que sería un "orden ideal". Como dice José Mourinho: "*los jugadores son talentos especiales (enfoque reduccionista), pero sin el equipo no pueden expresar todo lo que tienen, todo lo que son (enfoque sistémico). Por eso, todo se trata del equipo, de manera que un técnico no se debe entrenar jugadores de fútbol, sino equipos de fútbol*". De ahí que, el planteamiento táctico deba ser basado en las interacciones posibles entre los jugadores y lo que podría emerger como su resultado.

No obstante, a pesar de los planes tácticos, al ponerse en marcha el juego, las condiciones y los factores contextuales cambian. En este sentido, el orden *esperado* se contrasta con el orden *observado*, que puede cumplir o no las expectativas de la guía dada, pero que siempre dará información importante de lo como nuevas interacciones dan nuevos resultados y enriquecen el juego.

La autoorganización guiada en el partido también puede venir de las nuevas interacciones de jugadores determinantes que se comportan de manera sobresaliente en los momentos críticos (Clutch players). Muchos de ellos, tienen una mejor lectura y pueden generar interacciones de mayor peso (más relevantes) de las que dependerá en gran medida el futuro del juego-sistema.

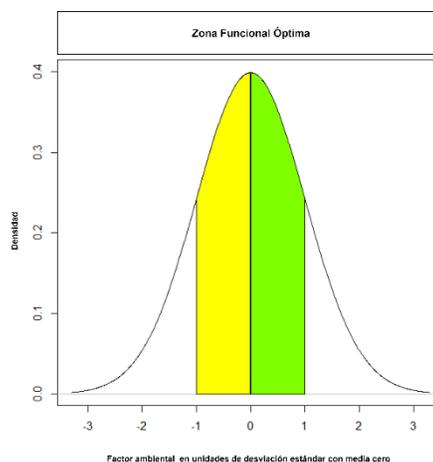

La gran cantidad de interacciones entre los elementos, hacen que aparezcan comportamientos, eventos o información nueva que no se esperaban (*emergencia*). Por ejemplo, de alguna manera podríamos saber todo de Messi, de Neymar y de Suárez por separado antes de jugar en el Barcelona, pero jamás hubiésemos podido imaginar las jugadas que emergieron al interactuar. MSN dio lugar a la *emergencia* de por lo menos 376 situaciones nuevas que derivaron en gol.

Las propiedades de autoorganización y emergencia descritas son las que hacen que el sistema funcione de forma óptima, mientras se adapta a los cambios del medio.

*Figura 3 Zona Funcional Óptima ante un Factor Ambiental dada la Tolerancia del Individuo*

Es allí donde observamos que existe una nueva propiedad que busca establecer balances de distintos tipos, que se define como *Homeostasis* o equilibrio dinámico (Fernández et al., 2014). La homeostasis le permite permanecer al sistema en una zona funcional optima mientras hace frente a "perturbaciones" (Fig. 3). En esta zona el individuo, o sistema, se mueve entre límites inferior y superior (en amarillo y verde) alrededor del promedio, donde se da el nivel más alto de tolerancia al factor ambiental o perturbación (Fernandez, 2015).

Desde nuestro enfoque, las características que componen la homeostasis son: (i) la *Resiliencia* o el tiempo que demora el sistema en regresar a su "Zona Funcional Óptima", después de que una perturbación le ha impactado. Por ejemplo, cuánto tiempo demora un jugador en volver a su esquema táctico y/o su rendimiento óptimo después de que su equipo recibió un gol. (ii) La *Antifragilidad* o la capacidad de un jugador para mejorar el desempeño al enfrentar las adversidades. Un tipo de respuesta evolutiva que se soporta en un mecanismo de adaptación positiva, en medio de perturbaciones fuertes, ambientes volátiles e impredecibles (Equihua et al., 2020; Pineda et al., 2018) .

Es de resaltar que, la *antifragilidad* (como otras propiedades de la complejidad) actúa de lo individual a lo colectivo (bottom-up information flow), y que de lo colectivo se retroalimenta a lo individual (top-down information flow). De tal manera que lo que hace un jugador contagia al equipo y el juego colectivo que se genera retroalimenta de vuelta al jugador.

La importancia de la resiliencia y antifragilidad en la toma de decisiones es conocida por abanderados del fútbol moderno como Víctor Orta. El director deportivo del Leeds opina que se debe ser resiliente y anti - frágil para poder actuar y conectar dos ámbitos: el de los dueños y el del campo de juego. Según Víctor, es la forma adecuada de generar confianza ya que la separación de estos dos ámbitos, años atrás, generó mucho conflicto.

En un nivel de detalle mayor, si bien *resiliencia* y *antifragilidad* pueden vincularse con la *Vulnerabilidad* y la *Amenaza* que definen el *Riesgo*, también están asociadas con la *Determinación o firmeza en el carácter* (Grit)*.* Es claro que, la determinación resulta en osadía, en valor, en consistencia y en solidez para triunfar en el juego, en los partidos y, como tal, en la vida.

El Grit viene de la sinergia entre la *Consistencia del Interés* (enfoque en intereses profesionales) y la *Perseverancia del Esfuerzo* (enfoque en el terreno de juego) (Duckworth et al., 2007). Ser consistente y perseverante tiene que ver mucho con lo que inicia todo: con los pensamientos, que son los que definen nuestro destino. Ya lo estableció Lao-Tse en su tiempo cuando dijo: siembra un pensamiento, cosecharás una acción; siembra una acción, cosecharás un hábito; siembra un hábito, cosecharás un carácter; siembra un carácter, cosecharás un destino.

Cabe destacar que los pensamientos, en el ámbito deportivo, han de ser administrados por una "mente entrenada-disciplinada" pues de no ser así, al hacer sinergia con las creencias[1] limitantes, el resultado podría ser más que frustrante. Por el contrario, si los pensamientos hacen sinergia con las creencias expansivas[2], el efecto aditivo y multiplicativo será el incremento en el desempeño profesional y el crecimiento en cuanto a desarrollo humano.

---

[1] Creencias que nos bloquean a la hora de conseguir nuestros objetivos. No tienen por qué ser reales, pero nos dificultan alcanzar aquello que realmente deseamos.

[2] Creencias que te hacen sentir bien y tener emociones placenteras. Las que te inspiran, te hacen mejorar como persona y te pueden llevar a alcanzar los retos más imposibles.

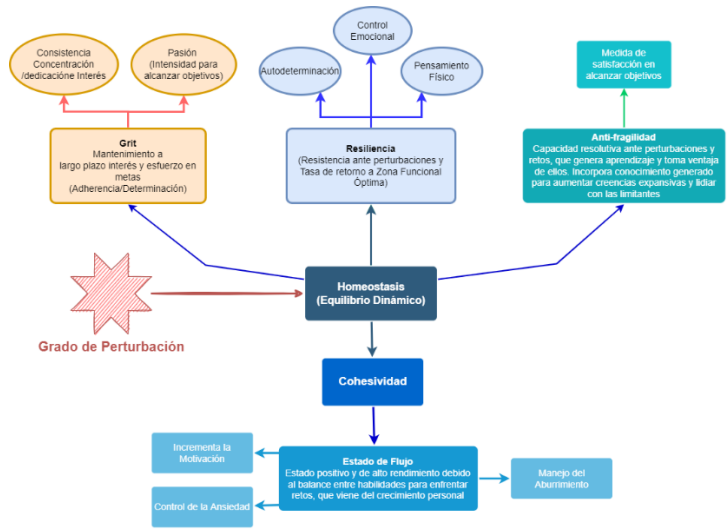

En este aspecto, el despliegue de las propiedades homeostáticas descritas se halla en los ámbitos de lo físico y lo cognitivo, por lo que tienen un sentido de integralidad y definen lo que podríamos estimar como *cohesividad* del jugador/equipo. Así se integra en la figura 4, que representa nuestro subsistema de valoración cognitivo.

Sobre la base de la *cohesividad* el jugador puede alcanzar estados de flujo apropiados (Csikszentmihalyi, 2014) donde sus habilidades cognitivas y físicas, relativas al juego, se emparejen con los retos que tenga lugar. En este momento si las habilidades son más grandes que los retos, el jugador entra en una zona de aburrimiento, por lo que se debe incrementar la dificultad del reto. Al incrementar el nivel de dificultad, sucederá que el reto sobrepasará las habilidades del jugador, quien entrará en una zona de estrés, pero que una vez superado por el desarrollo de nuevas habilidades fluirá al emparejar de nuevo las dos variables. En este momento el un rendimiento aumentado se consigue y se expresa como mejora continua.

*Figura 4 Variables e Indicadores del Subsistema Cognitivo de un Deportista/Equipo*

## 3. Cuando las cosas se mantienen igual: El peso del pasado y la gestión de cambio

Podemos destacar que, el enfoque científico de estudiar las partes ha resultado muy efectivo para sistemas que no cambian o cambian muy poco con el tiempo. Desde esta condición y desde una perspectiva analítica (análisis: acción de desligar, separar) será factible hallar algunas causas de lo sucedido en el presente desde la observación de los hechos pasados. Es por ello que podemos considerar que el equipo que ganó ayer, ganará hoy.

En el área de la predicción hay modelos considerados "ingenuos" (näive), que son muchas veces funcionales. Desde ellos, podemos decir que el mañana será como el hoy, dado que el hoy fue como el ayer.

Nada raro si consideramos que esta creencia se viene dando como resultado de un modelo de vida para muchos. Es decir, vivimos el presente explicándolo desde el pasado, y percibiendo que, el presente será probablemente una repetición del mismo (Fig. 6). En consecuencia, vemos que el futuro será, también con alta probabilidad, la repetición del presente. Y que, en relación al pasado, la vida se define como una redundancia cíclica. Será así siempre, a no ser que algo misterioso y mágico como una crisis profunda (la llamada noche oscura del alma) o "cisne negro" (un evento inesperado) nos saque de ese ciclo, el cual no es precisamente virtuoso.

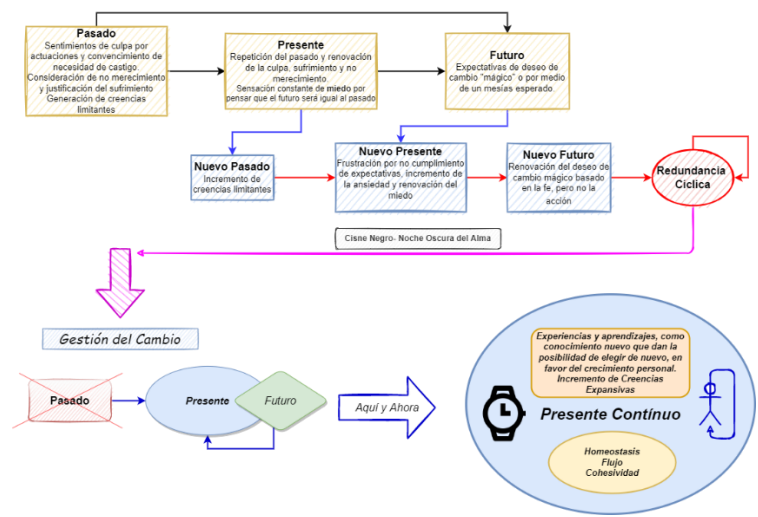

*Figura 5 Gestión del Cambio del Efecto del Tiempo*

En términos prácticos, la gestión del cambio debe tener lugar, principalmente para concebir todo momento como un presente continuo y no mirar hacia atrás. Lo que suceda en el transcurso de juego se verá como un reto que perfeccionará y dará más habilidades, como una oportunidad de aprender y decidir mejor desde la experiencia. Si se actúa con determinación y perseverancia en este presente continuo, el resultado siempre será el mejor.

## 4. Reconocer la complejidad es esencial para comprender el fútbol más allá

No reconocer al fútbol como un sistema complejo implica pasar por alto las interacciones relevantes, por lo que no podrán ser descifradas o entendidas en cuanto a su efecto. Habrá dificultad en identificar las propiedades asociadas de autoorganización, emergencia, homeostasis, entre otras posibles. Se generará cierta complicación para explicar muchos de los fenómenos que vemos dentro y fuera de la cancha. No podremos ver más allá del juego.

De no cambiar nuestros enfoques tradicionales que han pasado por alto la complejidad en el fútbol, seguramente tendremos en nuestro día a día la frustración de no ver el panorama en 360°.

Según Carlos Gershenson (2009), hay muchos factores adicionales que interactúan y afectan el futuro de un sistema, lo que lo hace complejo. Tal como ocurre en el fútbol en donde podríamos tener algún grado de acierto de cómo será la dinámica de un juego, no podremos tener la seguridad de su resultado hasta que no termine. Por ejemplo, cuando un equipo enfrenta a un rival superior de mayor jerarquía futbolística y económica, es razonable llegar a predecir que habrá más posibilidades de anotar un gol, pero será difícil tener el pronóstico acertado del juego.

En un entorno en el que se dan gran número de interacciones que expresan complejidad, se vale pensar también contraintuitivo dada la posibilidad de su resultado. Es por ello que algunos apostadores, no precisamente pesimistas, tienden a apostar en contra del favorito, por ser un resultado probable y de grandes dividendos económicos. Es observable que apostar en contra de lo obvio también desata gran cantidad de emociones en todos por la secreción de adrenalina, una consecuencia secundaria de la complejidad del juego.

## 5. Adaptación como respuesta a la complejidad

Como respuesta selectiva ante la imprecibilidad en el juego, ya sea que venga de la mentira que es el fútbol, o de su complejidad, existe la posibilidad de desarrollar la habilidad de cambiar el comportamiento ante los cambios que perturban el sistema. Es en este sentido que nos *adaptamos* con el fin de sobrevivir a las variaciones que percibimos. En la práctica, la adaptación también es requerida para afrontar muchos problemas que vienen de las "soluciones" que tomamos en el pasado y que ahora requieren de un nuevo cambio.

La adaptación, según lo expresa Stuart Kauffman, es un tipo de creatividad que va de la mano con la anticipación y la evolución ilimitada (Kauffman SA, 1991). Seguro podemos ver que debemos cambiar, de manera creativa, para no sufrir consecuencias indeseadas de una tendencia de cambio. En este aspecto actuamos de manera proactiva (antes que suceda) y hacemos del cambio mi oportunidad para mejorar. Lo contrario sería actuar de manera reactiva o actuar resolutivamente después que el cambio se ha dado para lidiar con sus efectos, lo que claramente es menos efectivo.

La adaptación y la proactividad hacen que siempre se quiera buscar una mejora substancial a algún proceso, lo que podría llevar a una innovación. Cabe resaltar que la innovación no necesariamente requiere conocimiento nuevo, más bien corresponde con evoluciones de los hechos o eventos desde sus estados anteriores. Es fácil reconocer que Whatsapp es un Messenger mejorado.

Para el caso del fútbol, las plataformas computacionales que hacen reconocimiento de video de distintas variables de desempeño de jugadores facilitan el proceso de visoria, aunque por ahora no las reemplaza al 100%. Tales plataformas basan su estructura en algoritmos de visión computacional que permiten el seguimiento de objetos. Se entiende entonces que son aplicaciones del conocimiento preexistente en las ciencias de la computación e inteligencia artificial y como innovaciones han mejorado al ser aplicadas al proceso de captar talento. Una innovación subsecuente, sobre esta misma base, sería la de desarrollar el talento, en particular, con la inclusión de la virtualidad y la realidad aumentada para la creación de contextos de entrenamientos particulares.

No saber adaptarse con la suficiente contribución, aporte y novedad a los retos que impone el fútbol, podría ser aún más traumático que las fallas en la predicción. En este sentido, habrá que valorar cuantas soluciones podemos dar ante un número indeterminado de posibles cambios. Es decir que, para no correr riesgos de subsistencia, debemos desarrollar una mayor variedad y riqueza de respuestas óptimas y eficaces. Aquí aplica la Ley de variedad de requerida que manifiesta que sólo la variedad absorbe variedad (Lambertz, 2020; Richards & Ashby, 1957). Esto nos permitirá afrontar un gran número de eventos y amenazas que podrían acontecer y explotar así nuestra *antifragilidad*.

En la práctica, el modelado y la simulación podrían ser de utilidad para anticipar escenarios futuros y generar información de utilidad para la toma de decisiones en términos de adaptación-anticipación. También la inteligencia artificial y la mecatrónica seguro tendrán cada vez más lugar en el fútbol.

Entre los retos que requieren adaptaciones desde la innovación están, otros temas: (i) ¿cómo mejorar el nivel y performance de los jugadores y árbitros?, (ii) ¿cómo especializar a los profesionales?, (iii) ¿cómo mejorar la experiencia de los aficionados?, (iv) desde las anteriores ¿cómo mejorar el nivel de los clubes y una liga?

La adaptación e innovación deben llegar a ser un activo de los clubes y parte de su modelo de negocio. Así que habría que pensar en secciones, departamentos o laboratorios de investigación y desarrollo, donde se haga innovación. Sin embargo, para ello, se requiere ya de un cambio cultural (que tarda, regularmente cuatro años). Como dice Simon Sinek: *la genialidad está en la idea y el impacto en la acción*, así que adaptarse en el fútbol debe ser un tema estructural en las agendas estratégicas y sus planes de implementación.

## 6. Dataísmo y Complejidad

Hoy en día, desde una perspectiva formal de ciencia de datos, y con el fin de estudiar las interacciones que expliquen la complejidad del fútbol, podríamos decir que, gracias a la explosión de mediciones, es viable estudiar "todas" las variables del juego. Así podríamos ver (i) ¿cuáles de ellas tienen mayor peso?, (ii) ¿cómo se relacionan entre sí? y (iii) ¿qué factores surgen de su aditividad o sinergia?

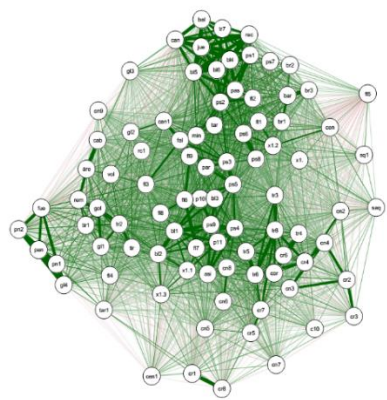

Tal como se observa en la figura 7 o red de correlación de las variables que caracterizan el desempeño de los jugadores de un equipo del fútbol profesional mexicano[3], existe una trama de asociaciones tanto directas como inversas de las que se puede partir en el estudio de las interrelaciones.

*Figura 6 Red de Correlación de Variables de desempeño de jugadores en el fútbol Mexicano*

Se presume que, desde la ciencia de datos, el entendimiento del fútbol podría tener muy buenos resultados, pues cada día tenemos más variables y más precisión, incluso con el seguimiento de los desplazamientos por GPS de los jugadores, que nos dan data hasta de las cargas musculares. En suma, podemos tener miles de variables de jugadores en distintas posiciones, así como de juegos por cada jornada. Adicionalmente, podemos analizar más de una temporada, bien sea para ver si hay tendencias o si hay diferencias entre algunas de ellas. Es decir, podemos hacer gran cantidad de ciencia con estos datos, principalmente porque no hay ciencia sin datos.

Pero… ¿por qué hay la percepción que tanto dato nos podría estar llevando a tener más dificultades en entender mejor el juego? ¿Podemos extraer, realmente, información útil para la toma de decisiones? O sólo hemos caído en lo que hemos llamado "*El Síndrome de los Datos Ricos e*

---

[3] Datos obtenidos de la plataforma Golstats

*Información Pobre-* (*Data Rich, Information Poor Syndrome-DRIPS*)"(Fernández et al., 2019), que nos ha convocado hoy a la urgente necesidad de reducir la dimensionalidad de los datos en factores y de generar indicadores que realmente lleven a mejorar la toma de decisiones.

¿Será por tal síndrome que los entrenadores y estrategas del fútbol confían más en su empirismo? O más aún, desde la perspectiva de la digitalización de la sociedad que plantea el filósofo de ascendencia coreana Byung-Chul Han (2022) ¿Estaremos a merced de un "dataísmo" descontextualizado y desprovisto de fundamentos? ¿Habremos caído en la "Infodemia", en los que tuits y memes nos muestran que no hay un verdadero entendimiento de los fenómenos deportivos ¿amén de los sociales? o ¿hemos caído en una "Infoburbuja" donde para reforzar mi identidad y mi trivialidad, sólo me informo de lo que me gusta y me alejo de ver la "Gran Fotografía" que es el juego? ¿Hacia dónde debemos mirar para inspirarnos y ser más creativos en las soluciones? ¿Qué podemos hacer si la dinámica de la producción de datos nos desborda?

Como opina Marco Garcés (director deportivo de los Ángeles Futbol Club[4]), éste es un eje crucial del fútbol; básicamente porque los datos pueden ayudar a tomar las decisiones, cuando las decisiones llevan tanto tiempo tomándose de forma subjetiva. Es allí donde surge uno de los mayores retos que se deben balancear, y consiste en cómo el analista de datos hace una correcta interpretación del fenómeno para producir información que sirva para la toma de decisiones, al tiempo que los tomadores de decisión, desde su juicio experto, le orientan en qué tipo de indicadores serían los apropiados. Se requiere por tanto un equilibrio entre lo cualitativo que ahonda en los rasgos y características, y lo cuantitativo que observa patrones y tendencias. También se requiere de un equilibrio entre el sentido teórico de las pruebas estadísticas y modelos, y lo que podría ser considerado la "tendencia" de moda entre los expertos (que podría no centrarse en el criterio adecuado que se podría tener).

Dado que a la fecha ambos lados de la ecuación (analista y juicio experto) tienen posibles sesgos, se sugiere un enfoque que involucre "el diálogo de saberes" en busca de una solución que integre las verdades que se dan de cada lado. Un reto más o como podríamos decir: una nueva piedra filosofal del fútbol dado lo multifactorial del tema. Misma que requiere de atención y profundidad para responderla científicamente, pero desde la base del empirismo que le ha dado la preponderancia al deporte.

Ante la riqueza de posibles soluciones existentes, no debemos descartar ahondar en la búsqueda y desarrollo de personas con habilidades especiales como los súper-pronosticadores. Una posibilidad que se desprende de lo relatado en el libro de Tetlock & Garner (2017), y que ha sido observada por directores deportivos como Víctor Orta del Leeds, como un hecho de gran interés. Víctor soporta su argumentación en situaciones como las vividas con su mentor "Monchi", de quien describe tenía habilidades especiales para pronosticar talento de manera impresionante. Particularmente cuando se le entregaba un listado predefinido sobre la base de un análisis exhaustivo de cualidades y calidades para seleccionar un jugador, como por ejemplo un lateral. Monchi les aseguraba: "el primero es el mejor lateral de todos. Pero el tercero es el mejor para el Sevilla ". El caso es que su pronóstico se cumplía. Esto muestra que, efectivamente existen personas con "un modo de pensar"

---

[4] https://www.lafc.com/news/marco-garces-info

(*cuestión de piel*) particular que les permite detectar talento y estimar la posible trayectoria de su carrera deportiva.

Son muchas las opciones que se tienen en este tipo de pronósticos y que se pueden adelantar como solución. Particularmente, en el campo de la computación y los sistemas expertos se abre la posibilidad de indagar en los "modelos mentales" y las reglas de decisión de quienes lo hacen con habilidad especial, para luego llevarlo a la inteligencia artificial como posible innovación.

Lo que observamos desde lo expuesto es que, existe la necesidad apremiante de tomar medidas en cuanto al dato, en "hacerlo hablar" para que sea realmente un soporte en las decisiones. No obstante, se requiere verificar tanto necesidades como soluciones en los equipos y el fútbol como un todo, y más allá de lo que pasa en la cancha.

Se requerirá por tanto tiempo, concertación y disposición para avanzar en los modelos de rendimiento adecuado, en el diseño de sistemas de valoración eficaces con indicadores hechos a la medida de cada dimensión del juego, y que tengan posibilidad de integración. Algo a considerar es que, al involucrar la dimensión científica se puede tardar en dar resultados, por lo que hay que pensar en implementar "metodologías ágiles". De ahí que resaltamos la necesidad de construir laboratorios de investigación y desarrollo dentro de los clubes, en los que como tarea fundamental y de base está el hacer las preguntas correctas/transformadoras.

## 7. La sencillez como base de la complejidad

Ante la realidad multidimensional del fútbol, que requiere ser reducida a factores determinantes sobre los cuales enfocarse y, dado que complejidad no es igual a complicado, algunas respuestas para enfrentar el entendimiento del juego como sistema complejo tienen que ver con explorar y buscar la sencillez en las soluciones o "El Arte de Hacerlo Simple". En tal sentido, de alguna forma tendremos que descomplicar lo que surge como complicado, pero que viene de lo simple.

Una alternativa puede ser tomar el camino de la practicidad donde menos es más y que igualmente genera complejidad. Según Luis Enrique[5] "al jugador se le debe dar *la mínima información* posible para que sea más efectivo. Se trata de dar información vital y una tarea exacta que pueda cumplir". Esta estrategia es similar a la distribución de tareas que vemos en las colonias de hormigas (Task allocation). Las hormigas tienen asignadas labores según su rol, al igual que en el fútbol. Están las obreras (equivalentes a volantes de recuperación), exploradoras (laterales), soldados (defensas), y la reina (delantero goleador).

Para cada rol existe una tarea principal y una secundaria de manera que, si se tapa la entrada de la colonia, no sólo las obreras se encargan de abrirla nuevamente, sino que las exploradoras cambian de rol y también ayudan. Este podría ser el caso de delanteros que también defienden o los defensas que van al ataque, obviamente, si es preciso y oportuno.

---

[5] t.ly/_XV2

El resultado de la sencillez no es la simplicidad del planteamiento y de la dinámica del juego, más que nada porque se genera una combinatoria de posibilidades (rol x número de acciones x situaciones asignadas x número de jugadores), que no es precisamente pequeña, en un contexto donde las nuevas interacciones son las que generan mayor riqueza de jugadas.

Algo importante para la asignación de tareas y su correcta ejecución por parte del jugador, es que el entrenador ejecute el *principio de no suponga, confié, pero verifique.*

El planteamiento de la simplicidad o "Descomplejizar lo Complejo" es una tesis de Miguel Álvarez[6], CEO de Kaantera (https://www.kaantera.com/) una plataforma de planificación y control de la actividad deportiva está tomando mucha fuerza. Miguel afirma que: "*No se trata de eliminar la complejidad, sino de ordenarla. Cuando generamos reglas de juego más fáciles, aceleramos la ejecución y eso genera una complejidad más efectiva*".

En este aspecto, lo que propone Miguel cabe dentro de lo que podría definirse como una extensión de la autoorganización guiada que produciría una "Complejidad Guiada". Es por ello que también Álvarez argumenta que: "*Al acelerar la complejidad, la dirigimos al objetivo que perseguimos*" y que "*esa complejidad es sencilla de entender en cómo se genera para el que la promueve, pero difícil de entender para el que se enfrenta a ella*". Este tipo de planteamientos es una muestra que, en el fútbol, en muchos visionarios existen los fundamentos tácticos para entender reglas e interacciones relevantes del sistema, por lo que el presente continuo, es más que prometedor para los disruptores de los métodos tradicionales y promotores de la complejidad.

## 8. Inteligencia vs Racionalidad

La inteligencia tiene como característica el tomar decisiones a corto plazo. Es evidente que, en el fútbol, estamos en un juego inteligente; sin embargo, de allí vienen muchas de las complicaciones que poseemos. La alternativa sería explorar la racionalidad, en busca de entender la complejidad del fútbol, pero la racionalidad requiere tiempo y la verdad: ya vienen el próximo partido… También deberíamos inspeccionar y ahondar más en lo cognitivo en cuanto al saber, la experiencia y el conocimiento que hemos generado, así como el conocimiento que deberíamos generar, pero la verdad: ya vienen las finales.

Sucede como cuando no sabemos cuál es la pregunta, pero estamos seguros que la respuesta es ¡23! Por ejemplo, el fútbol requiere de considerar muchas preguntas para llegar a ser, entre otros, un deporte de desarrollo humano. Pero, ¿cuáles son? Hay que considerar que siete palabras han dominado al fútbol: Siempre lo hemos hecho así (Anderson & Sally, 2013).

La reflexión es, si podemos armonizar dos dinámicas: la toma de decisiones inteligentes para el aquí y ahora, y la reflexión a mediano plazo para el entendimiento del fenómeno. Todo bajo un esquema de planeación estratégica y la aplicación de metodologías ágiles.

---

[6] https://www.linkedin.com/in/miguelalvarezrey/

De base, en el fútbol están todos los elementos para generar el conocimiento por distintas vías. Cada club, en sí mismo, es una universidad y una institución de investigación del más alto nivel. Verlo así, cambia el concepto del fútbol, de ser un deporte en el que se deben vivir emociones y diversión, a un deporte que, además de ello, tiene todos los elementos para la formación y desarrollo de un ser humano integral, mediado por la producción de conocimiento.

## 9. Información vs Anti-información

Desde la racionalidad se podría llegar a "ver más allá de lo evidente"; pensar en ¿qué no nos dice la información que hemos producido? Para ejemplificar, vale la pena pensar en lo sucedido en la segunda guerra mundial, como lo sugiere Carlos Gershenson[7] (investigador de la UNAM): "cuando los aviones llegaban a la base se hacía un análisis de los impactos recibidos con el fin de proteger esas zonas en el fuselaje". En cambio, alguien de manera contraintuitiva manifestó que esos puntos más que débiles eran fuertes, dado que finalmente los aviones habían llegado. Este postulado sugería que se debía observar zonas distintas a ellas, desde la pregunta: ¿Dónde habían sido impactados los aviones que no volvieron a la base?

Esta mecánica de cuestionamiento es viable de aplicar en el fútbol y por ello es apropiado comenzar a observar a los disruptores, a jugadores con habilidades especiales (superdotados), que regularmente no encajan en los esquemas tradicionales y que resuelven de manera particular. De alguna forma el fútbol se ha venido homogenizando como un *deporte de atletas de gran despliegue físico*, al tiempo que todos tienen la oportunidad de serlo desde la determinación. Entonces, ¿qué es lo que falta? ¿hacia dónde se debe mirar, que no se haya observado antes? ¿cuál es la anti-información que deberíamos inspeccionar?

## 10. Comentario final

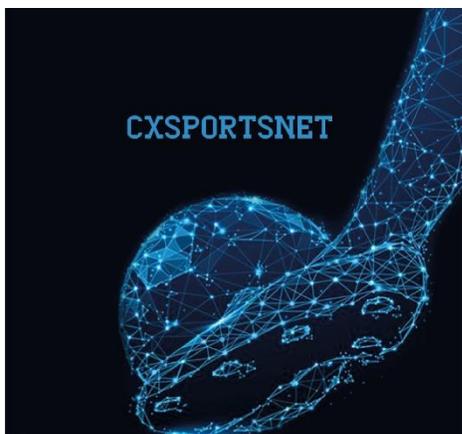

No ver la complejidad nos deja en el campo de las explicaciones limitadas. Si bien la verdad es contextual y todos tenemos una parte de ella, el enfoque de las ciencias de la complejidad da la oportunidad de integrar y ver la riqueza de aristas que tienen los problemas en el fútbol. De ahí mismo vendrá la riqueza de posibles soluciones. Si nos proponemos avanzar más en la comprensión del fútbol como sistema complejo tenemos más preguntas que respuestas que podrán ser resueltas desde el razonamiento y aprendizaje colectivo.

Así las cosas, hoy tenemos la necesidad de crear ese *día cero*, referido a la creación de nuevos contextos. El fútbol necesita continuar en la vía de la academia, es decir, generando conocimiento; por tanto, es momento de hacernos las preguntas adecuadas, las preguntas transformadoras. Es hora de escucharnos, y principalmente de generar redes de conocimiento. De esta manera, la idea de fútbol será un proyecto en construcción que creará un *nosotros*, que de seguro será más diverso y con gran riqueza de soluciones. Será un deporte complejo, pero no más difícil. De hecho, estamos convencidos que para ir más allá y solucionar lo que ha sido pasado por alto, se requiere de una nueva agenda para promover la cohesión de los actores y soluciones que sean realmente operativas, que faciliten realmente su trabajo como lo manifiesta Pol Llorente[8]. Es momento de actuar desde las interacciones relevantes, desde los datos, desde la experiencia que ha dado el empirismo, desde los aciertos de la heurística, desde la autoorganización, la emergencia, la homeostasis, la cohesividad, el flujo, la anti-información y, claramente, desde la adaptación como principio de complejidad.

---

[7] http://turing.iimas.unam.mx/~cgg/

## 11. Agradecimientos

## 12. Referencias

---

[8] https://www.rcdmallorca.es/primer-equipo/plantilla/mallorca/3
[9] https://twitter.com/chavaaguilera?lang=es